\documentclass[%
reprint,
 amsmath,amssymb,
 aps,
prb,
superscriptaddress
]{revtex4-2}

\usepackage{graphicx}
\usepackage{dcolumn}
\usepackage{bm}
\usepackage{mathrsfs}
\usepackage{braket}
\usepackage{color}
\usepackage{comment}
\usepackage{siunitx}

\usepackage{hyperref}
\hypersetup{
     colorlinks   = true,
     linkcolor    = blue,
     citecolor    = blue,
     urlcolor     = blue
}

\begin{document}

\preprint{APS/123-QED}

\title{Quantum geometric bounds for observables: \\ Linear responses, Drude weight, and orbital magnetization}

\author{Koki Shinada}
\altaffiliation[koki.shinada@riken.jp]{}
\affiliation{%
RIKEN Center for Emergent Matter Science (CEMS), Wako, Saitama 351-0198, Japan}
\author{Naoto Nagaosa}
\altaffiliation[nagaosa@riken.jp]{}
\affiliation{%
RIKEN Center for Emergent Matter Science (CEMS), Wako, Saitama 351-0198, Japan}%
\affiliation{Fundamental Quantum Science Program (FQSP), TRIP Headquarters, RIKEN, Wako 351-0198, Japan}

\date{\today}

\begin{abstract}
The quantum geometric tensor (QGT) provides nontrivial bounds among physical quantities, as exemplified by the metric-curvature inequality. 
In this paper, we investigate various bounds for different observables through certain generalizations of the QGT.
First, 
we demonstrate that bounds hold for all linear responses, which are produced by a QGT extended to many-body states, finite temperature, and general parameter space.
As an application, we show the thermodynamic inequality originating from the convexity of free energy can be further tightened.
Second, 
we establish a bound between the Drude weight and the orbital magnetization. 
The equality is exactly satisfied in the Landau level system, and systems with nearly flat bands tend to approach equality as well. 
We apply the resulting inequality to two orbital ferromagnets and support that the twisted bilayer graphene system is close to the Landau level system.
Moreover, we show that an analogous inequality also holds for a higher-order multipole, magnetic quadrupole.
Finally, we discuss the analogy between the QGT and the uncertainty principle, emphasizing that the existence of nontrivial bounds necessarily reflects quantum effects.
\end{abstract}

\maketitle

\section{Introduction}
Bounds expressed as inequalities provide fundamental insights in physics.
The second law of thermodynamics, which expresses the principle of entropy increase, and the uncertainty principle in quantum mechanics are representative examples, both constituting fundamental inequalities underpinning their respective theoretical frameworks.

The significance of bounds is increasingly being recognized in condensed matter physics as well.
In particular, these are closely related to the quantum geometry, describing the geometric structure of the Bloch wavefunctions in periodic crystals on a parameter space, such as the momentum space.
The bound between the quantum metric and the Berry curvature serves as a representative example \cite{PhysRevB.90.165139,peotta2015superfluidity,PhysRevB.104.045103}: the former is a Riemannian metric capturing local geometric structure, while the latter is a topological quantity characterizing global geometric features.
The bound highlights that the two geometric quantities mutually constrain each other, and the associated observable, the superfluid weight, is bounded \cite{peotta2015superfluidity}.
From a physical perspective, the former represents the spatial spread of the wavefunction in real space \cite{PhysRevB.56.12847,PhysRevLett.82.370}, while the latter reflects the non-commutativity of position operators in a crystal. This inequality can thus be regarded as an uncertainty principle in crystalline systems.
Moreover, in systems with strong orbital hybridization, such as topological insulators, the wavefunction is expected to spread more widely in real space \cite{PhysRevB.74.235111,PhysRevLett.98.046402}. The inequality is thus consistent with this physical expectation.
This localization length is also related to the band gap of insulators, and related inequalities have been actively studied \cite{PhysRevB.62.1666,PhysRevX.14.011052,PhysRevB.110.155107,PhysRevResearch.7.023158,10.21468/SciPostPhys.18.4.127}.
Furthermore, bounds involving other topological quantities have also been investigated \cite{PhysRevLett.124.167002,PhysRevB.107.L201106,PhysRevB.109.L161111,PhysRevB.111.L081103,yu2024universalwilsonloopbound,PhysRevB.110.075135,bouhon2023quantumgeometryprojectivesingle},
and another type of bound between the Drude weight and the quantum metric is studied \cite{PhysRevB.108.L140503}.

Inequalities themselves are essential relations by providing limitations to the magnitude of physical quantities; however, the conditions for equality are equally critical.
The aforementioned metric–curvature inequality attains equality in the Landau level system \cite{PhysRevB.104.045103}. Accordingly, the condition for equality, together with the uniformity of the Berry curvature in momentum space, provides a criterion for assessing the similarity to the Landau level system \cite{PhysRevB.90.165139,jackson2015geometric,PhysRevLett.114.236802,PhysRevB.96.165150,PhysRevB.104.045103,PhysRevLett.127.246403}.
Furthermore, the condition for equality, referred to as vortexability \cite{PhysRevResearch.2.023237,PhysRevB.108.205144,PhysRevResearch.5.023166,PhysRevLett.134.106502}, constitutes a sufficient condition for realizing fractional quantum Hall states (FQHS) in flat-band systems. Notably, it does not require uniform Berry curvature, indicating that FQHS can be formed even in systems that are not exact analogs of the Landau level system.

In this way, useful discussions on inequalities are being explored from the perspective of quantum geometry.
In particular, the quantum geometric tensor (QGT) plays a crucial role in the metric-curvature bound, which originates from the semipositive definiteness of the QGT.
In recent years, generalizations of QGT have been explored.
For instance, the time-dependent QGT is connected to certain sum rules associated with the electrical conductivity tensor \cite{komissarov2024quantum,verma2025instantaneousresponsequantumgeometry,Verma_2025},
and an extension to correlated systems using Green's function method with a self-energy correction is discussed \cite{PhysRevB.107.125116}.
In addition, a QGT that connects the Drude weight and the orbital magnetization has been proposed \cite{resta2017}.
Furthermore, the extension to higher-order geometric quantities such as connection and its relation with nonlinear responses have recently been discussed \cite{ahn2022riemannian,PhysRevLett.133.186601,mitscherling2025gaugeinvariantprojectorcalculusquantum,avdoshkin2025multistategeometryshiftcurrent}.
An important question is how far the QGT can be generalized, what physical quantities it can encompass, and what kinds of constraints it imposes on the physical quantities it describes.

In this paper, we discuss several aspects of the generalization of the QGT. Furthermore, we identify the physical quantities described by the QGT and examine the bounds among them.
In Sec.~\ref{sec_linear}, 
we demonstrate that a general linear response function produces an extended QGT to many-body states, finite temperature, and general parameter space and that it gives bounds for all linear responses.
We present some applications and discuss an inequality that is tighter than the thermodynamic bound as an example.
In Sec.~\ref{sec_orbital}, 
we provide a detailed analysis of the bound between the Drude weight and orbital magnetization using a generalized QGT, and we analyze the condition of equality using some models.
We further show that an analogous bound holds for a higher-order multipole, spin magnetic quadrupole. Moreover, we elucidate the connection of this bound to the uncertainty principle, highlighting the essential role of quantum effects.

\section{bounds in linear response theory} \label{sec_linear}
\subsection{Proof of inequalities}
We consider a system described by a Hamiltonian $\hat{H}$ with a (discrete) translation symmetry and a time translation symmetry.
Here, we consider a general Hamiltonian without assuming whether the system consists of fermions or bosons, whether it is interacting or non-interacting. 
Furthermore, we do not assume whether the system is at zero temperature or finite temperature.
Then, a general linear response function is written, using the Lehmann representation with a wavenumber $\bm{q}$, as
\begin{equation}
    \Phi_{O^i O^j}
    =
    -\frac{1}{V} \sum_{nm} \frac{\rho_{nm}}{\tilde{\omega} + E_{nm}} O^i_{nm}(\bm{q}) O^j_{mn}(-\bm{q}).
\end{equation}
Here, $\rho_n = e^{-\beta(E_n - \mu N_n)}/\Xi$ is the grand canonical distribution with the energy $E_n$ and the particle number $N_n$ for a microstate $\ket{n}$, and $\Xi = \sum_n e^{-\beta(E_n - \mu N_n)}$ is the grand partition function.
$\mu$ and $\beta$ are a chemical potential and an inverse temperature, respectively.
$O^i_{nm}(\bm{q}) = \braket{n | \hat{O}^i(\bm{q}) | m}$ is the transition matrix of a second-quantized operator $\hat{O}^i(\bm{q}) = \int_V d\bm{r} \hat{O}^i(\bm{r}) e^{i \bm{q} \cdot \bm{r}}$. Because a local operator $\hat{O}^i(\bm{r})$ is Hermitian, the identity $\hat{O}^i(\bm{q})^{\dagger} = \hat{O}^i(-\bm{q})$ is satisfied. 
In addition, we impose the commutativity with the particle number operator on the operator $\hat{O}^i(\bm{r})$.
In this case, the particle number is not changed by the operator $\hat{O}^i(\bm{r})$, and thus, we get $N_n = N_m$.
$V$ is the volume of the system.
We use some abbreviations: $\rho_{nm} = \rho_n - \rho_m$, $E_{nm} = E_n - E_m$, and $\tilde{\omega} = \omega + i\delta$, where $\delta = +0$ is the adiabatic factor.

The absorptive part of this tensor is given by
\begin{align}
    \Phi^{\mathrm{abs.}}_{O^iO^j} &= \frac{1}{2 i} [\Phi_{O^iO^j} - (\Phi_{O^jO^i})^*] \nonumber \\
    &=
    \mathrm{Im} \Phi^{\mathrm{(S)}}_{O^iO^j} - i \mathrm{Re} \Phi^{(\mathrm{AS})}_{O^iO^j} \nonumber \\
    &=
    \frac{\pi}{ V} \sum_{nm} \rho_{nm} O^i_{nm}(\bm{q}) O^j_{mn}(-\bm{q}) \delta(\omega + E_{nm}).
\end{align}
This tensor is Hermitian by definition.
(S) and (AS) are the symmetric part and antisymmetric part for the interchange of indices $i$, $j$.
In the regime of non-negative frequency $\omega \geq 0$, this part has the semipositive definiteness, i.e., its eigenvalues are semipositive, as shown in the following.
There, $E_{nm} \leq 0$ leads to $\rho_{nm} \geq 0$, and for any complex vector $\bm{v}$, this tensor satisfies
\begin{equation}
v_i^{*} \Phi^{\mathrm{abs.}}_{O^i O^j} v_j
=
\frac{\pi}{V} \sum_{nm} \rho_{nm} |\tilde{v}|^2 \delta(\omega + E_{nm}) \geq 0.
\end{equation}
Here, we define a complex value $\tilde{v} = O^j_{mn}(-\bm{q}) v_j$.
The integral of this tensor divided by $\omega^{\alpha}$ over the range $\omega=+0$ to $\infty$ 
\begin{align}  \label{def_M}
   \int^{\infty}_0 \mathscr{P} \frac{\Phi^{\mathrm{abs.}}_{O^iO^j}}{\omega^{\alpha}} d\omega
   =&
   \frac{\pi}{V} \sum_{E_n < E_m} \frac{\rho_{nm}}{(E_{mn})^{\alpha}} O^i_{nm}(\bm{q}) O^j_{mn}(-\bm{q}) \nonumber \\
   \equiv&
   M^{\alpha}_{O^iO^j}(\bm{q})
\end{align}
also satisfies semipositive definiteness.
$\alpha$ is taken to be an integer in the following discussion.
$\mathscr{P}$ denotes the principal value of the integral, meaning that $\omega = 0$ is excluded from the integration range.
This integral yields one of the sum rules of linear responses.
We note that the semipositive definiteness is maintained even if the frequency integral is restricted to a certain range of non-negative frequencies.

This semipositive definiteness leads to some nontrivial inequalities between the real and imaginary parts of this tensor $M^{\alpha}_{O^i O^j}(\bm{q})$.
Because eigenvalues of the semipositive matrix are not negative, resulting in $\mathrm{det} M^{\alpha}(\bm{q}) \geq 0$, 
if we choose two operators $\hat{O}^1$ and $\hat{O}^2$ and construct a $2 \times 2$ matrix composed of these operators, 
the semipositive definiteness gives an inequality
\begin{align}
    \sqrt{\mathrm{det}[\mathrm{Re} M^{\alpha}(\bm{q}) ]} \geq |\mathrm{Im} M^{\alpha}_{O^1O^2}(\bm{q})|.
\end{align}
Furthermore, the real part itself is semipositive, thus the inequality of arithmetic and geometric means gives
\begin{equation}
    \frac{1}{2} \mathrm{tr} [ \mathrm{Re} M^{\alpha}(\bm{q}) ] \geq \sqrt{ \mathrm{det} [ \mathrm{Re} M^{\alpha}(\bm{q}) ] }.
\end{equation}
These inequalities provide bounds between various linear responses.
We list several candidate operators in TABLE.~\ref{table_conjugate}.
\begin{table}[t] 
    \caption{List of the relation between external field and its conjugate operator.}
    \label{table_conjugate}
    \renewcommand{\arraystretch}{1.3}
    \begin{tabular}{@{\hskip 8pt}l@{\hskip 15pt}l@{\hskip 8pt}}
    \hline
        External field & Conjugate operator \\ \hline \hline
        vector potential $A$  & electric current $\hat{J}$ \\
        scalar potential $\phi$ & electric density $\hat{n}$ \\
        electric field $E$ & electric polarization $\hat{P}_e$ \\
        magnetic field $B$ & spin $\hat{S}$ \\
        strain tensor $\partial u$ & stress tensor $\hat{\sigma}$ \\
        \hline
    \end{tabular}
\end{table}
In this way, the semipositive definiteness and the resulting inequalities are general properties inherent to the linear response function.

Finally, we comment on the constraints imposed by time-reversal symmetry.
Onsager reciprocal theorem states
\begin{equation}
    \Phi_{O^i O^j}(\omega, \bm{q}, \bm{B}) = \varepsilon_{i} \varepsilon_{j} \Phi_{O^j O^i}(\omega, -\bm{q}, -\bm{B}),
\end{equation}
where $\varepsilon_i (= \pm 1)$ is the sign determined by the time-reversal operation $\hat{\mathcal{T}}$ and is defined by $\hat{\mathcal{T}} \hat{O}^i(\bm{r}) \hat{\mathcal{T}}^{-1} = \varepsilon_i \hat{O}^i(\bm{r})$.
$\bm{B}$ is a time-reversal symmetry ($\mathcal{T}$-symmetry) breaking term, such as a magnetic field.
Therefore, either the real or the imaginary part of the off-diagonal parts at the limit $\bm{q} \to 0$, $M^{\alpha}_{O^1 O^2}(0)$, becomes zero when $\mathcal{T}$-symmetry is preserved.

\subsection{Relation with quantum geometric tensor} \label{sec_QGT}
Considering the uniform limit $\bm{q} \to 0$ at zero temperature
, the tensor $M^{\alpha}_{O^i O^j}$ is related to a quantum geometric tensor (QGT) for a many-body ground state on a general parameter space.
Here, the many-body ground state at zero temperature $\ket{\mathrm{GS}} \equiv \ket{n = 0}$ is defined as a microstate satisfying the condition $E_0 - \mu N_0 < 0$, i.e., $\rho_0 = 1$, and it is assumed to be nondegenerate. Excited states $\ket{n \geq 1}$ are defined as microstates satisfying the condition $E_n - \mu N_n > 0$, i.e., $\rho_n =0$.
We introduce a parameter $X_i$, which is conjugate to the operator $\hat{O}^i$. 
It is included in a Hamiltonian by $X_i \hat{O}^i$.
Then, using the Hellmann–Feynman theorem, where $\braket{n | \partial \hat{H} / \partial X | m} = - E_{nm} \braket{ n | \partial_X m } $ for $n \neq m$,
the tensor can be rewritten by a QGT as
\begin{equation} \label{GS_QGT}
    M^{\alpha}_{O^iO^j}
    =
    \frac{\pi}{V} \braket{ \partial_{X_i} 0 |  \hat{Q} ( \hat{H} - E_0 )^{2-\alpha} \hat{Q} | \partial_{X_j} 0 }.
\end{equation}
Here, $\hat{Q} = 1- \hat{P}$, where $\hat{P} = \ket{0} \bra{0}$ is a projection operator to the ground state, and we assume $\braket{ 0 | \hat{Q} | 0 }$ = 0.
From this expression, we can conclude that the sum rule of the absorptive part of the linear response function constitutes an extension of the QGT to rather general situations, including many-body systems, finite temperature, and general parameter space.

\subsection{Thermodynamic inequality and tighter relation}
As an application of inequalities derived above, we consider the case of $\alpha =1$ and the uniform limit $\bm{q} \to 0$.
In this case, the real part of the tensor $M^1_{O^i O^j}(0)$ corresponds to a static susceptibility $\chi_{O^iO^j}$,
\begin{align}
    \mathrm{Re} M^1_{O^i O^j}(0) &= -\frac{\pi}{V} \sum_{E_n < E_m} \frac{\rho_{nm}}{E_{nm}} \mathrm{Re} [O^i_{nm} O^j_{mn}] \nonumber \\
    &\equiv \pi \chi_{O^iO^j} /2. \label{M1_OO_static}
\end{align}
For instance, when the operator $\hat{O}^i$ is a spin operator or an electric polarization operator, the above equation is a static spin susceptibility $\chi^{\mathrm{m}}_{ij} = - \partial^2 \Omega/ \partial h_i \partial h_j$ or an electric polarizability $\chi^{\mathrm{e}}_{ij} = - \partial^2 \Omega/ \partial E_i \partial E_j$, respectively.
Here, $\Omega [\beta , \mu, X] = - (\beta V)^{-1} \ln \Xi[\beta , \mu, X]$ is a grand potential density.
Note that the tensor $M^{\alpha}_{O^i O^j}$ is defined for the absorptive part of the linear response function, while $\chi_{O^iO^j}$ is a static susceptibility. They are related by the Kramers-Kronig relation (see Appendix \ref{appen_KK} for details).

For general $\alpha$, the real part $\mathrm{Re} M^{\alpha}$ itself is semipositive, thus $\det \mathrm{Re} M^{\alpha} \geq 0$.
In the present case, this inequality becomes a thermodynamic inequality originating from the convexity of the grand potential,
\begin{equation}
    \chi_{O^1 O^1} \chi_{O^2 O^2} - ( \chi_{O^1 O^2} )^2 \geq 0.
\end{equation}
However, we note that, due to the presence of the imaginary part, a tighter inequality can be obtained compared to this thermodynamic inequality,
\begin{equation} \label{tight_ineq}
    \chi_{O^1 O^1} \chi_{O^2 O^2} - ( \chi_{O^1 O^2} )^2 \geq (2 \mathrm{Im} M^1_{O^1 O^2}(0)  / \pi)^2.
\end{equation}

In the following, we will discuss two examples.
When we set spin operators $\hat{S}^{1,2}$ (here, the indices 1 and 2 denote one of the real-space coordinates, $x$, $y$, or $z$. The same applies to the following) as the operators $\hat{O}^{1,2}$,
we obtain an inequality
\begin{equation} \label{M1_SS}
    \frac{\pi}{2} \sqrt{ \chi^{\mathrm{m}}_{11} \chi^{\mathrm{m}}_{22} - (\chi^{\mathrm{m}}_{12})^2 } \geq \Bigl| [ \mathrm{Re} \chi^{\mathrm{m}(\mathrm{AS})}_{12}]^1 \Bigr|.
\end{equation}
Here, we use an abbreviation about a sum rule: $[A]^{\beta} = \int^{\infty}_0 A(\omega) / \omega^{\beta} d\omega$.
The real part is a spin susceptibility, as introduced above. The imaginary part is the spectral summation of the antisymmetric part of the dynamical spin susceptibility $\chi^{\mathrm{m}}(\omega)$.
As discussed above, the imaginary part is finite only when $\mathcal{T}$-symmetry is broken, and it leads to the tighter inequality in Eq.~(\ref{tight_ineq}).
If $\mathcal{T}$-symmetry is not broken, the tighter inequality is reduced to the thermodynamic inequality.
If $\varepsilon_1 = \varepsilon_2$, the same discussion is valid for other linear response functions.

For the other example, we set a polarization operator $\hat{P}$ and a spin operator $\hat{S}$ as the operators $\hat{O}^{1,2}$, respectively.
Here, although subscripts for $\hat{P}$ and $\hat{S}$ are omitted, one of the components along $x$, $y$, or $z$ is implicitly selected. The same applies below.
In this case, $\varepsilon_1 = - \varepsilon_2$,
and the resulting inequality is
\begin{equation} \label{M1_PS}
    \frac{\pi}{2} \sqrt{ \chi^{\mathrm{e}} \chi^{\mathrm{m}} - ( \chi^{\mathrm{me(S)}} )^2  } \geq \Bigl| [ \mathrm{Re} \chi^{\mathrm{me}(\mathrm{AS})}]^1 \Bigr|.
\end{equation}
Here, $\chi^{\mathrm{me(S)}}$ is a static $\mathcal{T}$-odd magnetoelectric polarizability, and $\chi^{\mathrm{me(AS)}}(\omega)$ is a $\mathcal{T}$-even optical magnetoelectric response function, which can be observed by optical activity.
While the thermodynamic bound on the magnetoelectric susceptibility derived from the inequality for the real part is well known \cite{PhysRev.168.574}, here a tighter inequality is obtained due to the contribution of the imaginary part.

\subsection{Some examples of inequality}
We will discuss some examples of inequalities.
However, inequalities for all other linear responses and cross correlations are established and can be produced in the same manner as discussed below.

\subsubsection{\texorpdfstring{Case: $\hat{O}^1 = \hat{J}^{1}$ and $\hat{O}^2 = \hat{J}^{2}$}{Case1}}
We choose the $i$-th component of the electric current operator $\hat{J}^i$ as the operator $\hat{O}^i$.
In the case of $\alpha = 1$, the tensor corresponds to the spectral summation of optical conductivity $\sigma_{ij}(\omega)$. 
The resulting inequality is
\begin{equation} \label{M1_JJ}
    \frac{1}{8\pi} \sqrt{ \omega^2_{\mathrm{p},11} \omega^2_{\mathrm{p},22} -\omega^4_{\mathrm{p},12} } \geq \Bigl| [ \mathrm{Im} \sigma^{(\mathrm{AS})}_{12} ]^0  \Bigr|.
\end{equation}
Here, $\omega_{\mathrm{p},ij}$ is the plasma frequency matrix, and $\omega_{\mathrm{p},ij} = \sqrt{4 \pi e^2 n_e / m_e} \delta_{ij}$ for a non-relativistic Hamiltonian, where $n_e$ is a number density of electron and $m_e$ is an electron mass.
The imaginary part is the sum rule of the magnetic circular dichroism.

In the case of $\alpha =3$, the real part is a static electric polarizability $\chi^{\mathrm{e}}_{ij}$ and the imaginary part is the spectral summation of the optical Hall conductivity divided by $\omega^2$.
The resulting inequality is
\begin{equation} \label{M3_JJ}
    \frac{\pi}{2} \sqrt{ \chi^{\mathrm{e}}_{11} \chi^{\mathrm{e}}_{22} - (\chi^{\mathrm{e}}_{12})^2 } \geq \Bigl| [\mathrm{Im} \sigma^{(\mathrm{AS})}_{12} ]^2 \Bigr|.
\end{equation}
These inequalities can be derived by using the time-dependent QGT for noninteracting systems \cite{komissarov2024quantum,verma2025instantaneousresponsequantumgeometry,Verma_2025}.

\subsubsection{\texorpdfstring{Case: $\hat{O}^1 = \hat{J}$ and $\hat{O}^2 = \hat{S}$}{Case3}}
We consider a cross correlation between electricity and magnetism. 
We choose an electric current operator and a spin operator as $\hat{O}^i$.
In the case of  $\alpha =2$, the resulting inequality is
\begin{equation}
    \sqrt{[\mathrm{Re}\sigma^{(\mathrm{S})} ]^1 [\mathrm{Im} \chi^{\mathrm{m}(\mathrm{S})}]^2 - ([\mathrm{Re} \chi^{\mathrm{me}(\mathrm{AS})}]^1)^2 } \geq \frac{\pi}{2} | \chi^{\mathrm{me}(\mathrm{S})} |.
\end{equation}
Here, we obtain another different inequality for $\chi^{\mathrm{me}(\mathrm{S})}$ from the thermodynamic inequality in Eq.~(\ref{M1_PS}).

Other linear response functions also give various inequalities in the same manner.
We list several candidates in TABLE.~\ref{table_linear}.
\begin{table*}[t]
\caption{List of linear response functions corresponding to the tensor for $\alpha =1$ ($M^1_{O^i O^j}$). The listed symbols are defined as follows:
$\omega_{\mathrm{p}}$ is the plasma frequency, $\sigma(\omega)$ is the optical conductivity, $\sigma_{\mathrm{Hall}}$ is the DC Hall conductivity, $\chi^{\mathrm{e}}$ is the static electric polarizability, $\chi^{\mathrm{me}}$ is the static magnetoelectric polarizability, $\chi^{\mathrm{me}}(\omega)$ is the optical magnetoelectric responses, $\chi^{\mathrm{m}}$ is the static spin susceptibility, $\chi^{\mathrm{m}}(\omega)$ is the dynamical spin susceptibility, $Z$ is the Born effective charge, $Z(\omega)$ is the dynamical Born effective charge, $d$ is the piezoelectric coefficient, $Q$ is the piezomagnetic effect, $\lambda$ is the elastic modulus tensor, $d(\omega)$ is the dynamical piezoelectric effect, $Q(\omega)$ is the dynamical magnetopiezo effect, $\lambda(\omega)$ is the dynamical elastic modulus tensor.
Some responses have no established names, however, we name them as follows:
$\chi_{Au/A \partial u}$ is the static current induced by displacement and strain,
$\chi_{Au/A \partial u}(\omega)$ is the dynamical one,
$Z^{\mathrm{m}}$ is the Born effective magnetization, which means a magnetization driven by displacement of atoms,
$Z^{\mathrm{u}}$ is the static displacement susceptibility,
$Z^{\mathrm{m/u}}(\omega)$ is the optical version of them,
$\chi_{u\partial u}$ is the static displacement induced by strain, $\chi_{u\partial u}(\omega)$ is its dynamical one.
We use an abbreviation about a sum rule: $[A]^{\beta} = \int^{\infty}_0 A(\omega) / \omega^{\beta} d\omega$.
(S) and (AS) denote the symmetric and antisymmetric parts under the exchange of indices. 
Each cell in the table displays two physical quantities: the real part of the tensor $M^1_{O^iO^j}$ on the left and its imaginary part on the right.
The real part can be rewritten in terms of the static response function via the Kramers–Kronig relation (see Appendix \ref{appen_KK} for details).
}
\label{table_linear}
\renewcommand{\arraystretch}{1.5}
\begin{tabular}{@{\hskip 15pt}c@{\hskip 15pt}||@{\hskip 15pt}c@{\hskip 15pt}c@{\hskip 15pt}c@{\hskip 15pt}c@{\hskip 15pt}c@{\hskip 15pt}} \hline
        & $A$ & $E$ & $B$ & $u$ & $\partial u$  \\ \hline \hline
        $A$  & $\omega_{\mathrm{p}}$, $[ \mathrm{Im} \sigma^{(\mathrm{AS})} ]^0$ & $\sigma_{\mathrm{Hall}}~\mathrm{or}~0$, $[ \mathrm{Re} \sigma^{\mathrm{(S)}}]^1$ &  $0$, $[ \mathrm{Im} \chi^{\mathrm{me(S)}}]^0$   & $\chi_{Au}$, $[\chi_{Au}^{\mathrm{(AS)}}]^1$ &  $\chi_{A \partial u}$, $[\chi_{A \partial u}^{\mathrm{(AS)}}]^1$  \\
        $E$ & & $\chi^{\mathrm{e}}$, $[ \mathrm{Im} \sigma^{(\mathrm{AS})} ]^2$ & $\chi^{\mathrm{me(S)}}$, $[ \mathrm{Re} \chi^{\mathrm{me}(\mathrm{AS})}]^1$ & $Z, [\mathrm{Re}Z^{(\mathrm{AS})}]^1$ & $d$, $[\mathrm{Re}d^{\mathrm{(AS)}}]^1$ \\ 
        $B$ & & & $\chi^{\mathrm{m}}$, $[ \mathrm{Re} \chi^{\mathrm{m}(\mathrm{AS})}]^1$ &  $Z^{\mathrm{m}}$, $[\mathrm{Re}Z^{\mathrm{m(AS)}}]^1$ & $Q$, $[\mathrm{Re} Q^{\mathrm{(AS)}}]^1$  \\
        $u$ &  & & & $Z^{\mathrm{u}}$, $[\mathrm{Re}Z^{\mathrm{u(AS)}}]^1$ & $\chi_{u\partial u}$, $[\chi_{u\partial u}^{(\mathrm{AS})}]^1$ \\
        $\partial u$ & & & & & $\lambda$, $[\mathrm{Re} \lambda^{\mathrm{(AS)}}]^1$ \\ 
\hline
\end{tabular}
\end{table*}

\subsection{Noninteracting periodic system}
In the following, we focus on noninteracting periodic and fermionic systems described by the Bloch states.
Throughout the following, all operators are represented in the first-quantization.
In this case, the tensor $M^{\alpha}_{O^i O^j}(\bm{q})$ reads
\begin{equation}
    M^{\alpha}_{O^iO^j}(\bm{q})
    =
    \pi \int_{\bm{k}} \sum_{\substack{\epsilon_{n\bm{k}+\bm{q}/2} \\ < \epsilon_{m\bm{k}-\bm{q}/2}}} \frac{f_{nm\bm{k},\bm{q}}}{(\epsilon_{mn\bm{k},\bm{q}})^{\alpha}} O^i_{nm}(\bm{q}) O^j_{mn}(-\bm{q}),
\end{equation}
where $O^i_{nm}(\bm{q}) = \braket{u_{n\bm{k} + \bm{q}/2} | \hat{O}^i  |u_{m\bm{k} - \bm{q}/2} }$, $f_{nm\bm{k},\bm{q}} = f_{n\bm{k}+ \bm{q}/2} - f_{m\bm{k}-\bm{q}/2}$, $\epsilon_{mn\bm{k},\bm{q}} =  \epsilon_{m\bm{k}-\bm{q}/2} - \epsilon_{n\bm{k}+\bm{q}/2}$, and $\int_{\bm{k}} = \int_{\mathrm{BZ}} d^d\bm{k}/(2\pi)^d$.
The notations introduced above are summarized below.
The Bloch Hamiltonian $\hat{H}_{\bm{k}}$ with a Bloch wavenumber $\bm{k}$ is diagonalized by the periodic part of a Bloch function $\ket{u_{n\bm{k}}}$ with a band index $n$, and the eigenenergy is $\epsilon_{n\bm{k}}$.
$f_{n\bm{k}} = 1/(e^{\beta (\epsilon_{n\bm{k}} - \mu ) } + 1)$ is the Fermi distribution function.

In the following discussion, we take the limit $\bm{q} \to 0$.
When taking the limit, we need to be careful about the intraband contribution ($n = m$), because the energy difference $\epsilon_{nn} (= 0)$ in the denominator can potentially cause a divergence. 
In the case of $\alpha \geq 2$, this contribution causes a divergence.
Accordingly, to exclude this divergence, we consider only the interband contribution.
In the case of $\alpha \leq 1$, no singularities are caused. 
In particular, the intraband contribution makes the Fermi surface term $\partial f_{n\bm{k}} / \partial \epsilon_{n\bm{k}}$ appear with the factor $1/2$ in the case of $\alpha =1$, and it becomes zero for other cases.
This factor of $1/2$ arises because the contribution from the Fermi surface appears as an absorption spectrum at $\omega \to 0$, and only half of it contributes due to the integration being restricted to the positive frequency domain.
These remarks are summarized in the TABLE~\ref{table_contribution_M}.
We note that, for insulating systems with a finite excitation gap, the divergence problem is absent, and the tensor $M^{\alpha}_{O^iO^j}$ can be fully expressed by the interband contribution.
\begin{table}[t] 
    \caption{The contribution to the tensor $M^{\alpha}_{O^i O^j}(\bm{q} = 0)$ for noninteracting cases. Checkmark (\checkmark) means ``included".}
    \label{table_contribution_M}
    \renewcommand{\arraystretch}{1.3}
    \begin{tabular}{@{\hskip 8pt}c@{\hskip 8pt}|@{\hskip 8pt}c@{\hskip 8pt}c@{\hskip 8pt}}
    \hline
          & intraband & interband \\ \hline \hline
        $\alpha \geq 2$ &  excluded  & \checkmark \\
        $\alpha =1$ & Fermi surface term & \checkmark \\
        $\alpha \leq 0$  & zero & \checkmark \\ \hline
    \end{tabular}
\end{table}

The interband contribution to the tensor $M^{\alpha}_{O^i O^j}$ is transformed to a quantum geometric tensor on a general parameter space at zero temperature.
The interband contribution $\tilde{M}^{\alpha}_{O^i O^j}$ is
\begin{equation}
    \tilde{M}^{\alpha}_{O^i O^j} = \pi \int_{\bm{k}} \sum_{n:\mathrm{occ}} \braket{ \partial_{X_i} u_{n\bm{k}} | \hat{Q}_{\bm{k}} ( \hat{H}_{\bm{k}} - \epsilon_{n\bm{k}} )^{2-\alpha} \hat{Q}_{\bm{k}} | \partial_{X_j} u_{n\bm{k}} }. \label{QGT_linear_response}
\end{equation}
Here, we define $\hat{Q}_{\bm{k}} = 1 - \hat{P}_{\bm{k}}$, using a projection operator to occupied states $\hat{P}_{\bm{k}} = \sum_{n:\mathrm{occ}} \ket{u_{n\bm{k}}} \bra{u_{n\bm{k}}}$.
This tensor for noninteracting systems reproduces the well-known metric-curvature inequality \cite{PhysRevB.90.165139,peotta2015superfluidity}.
We choose an electric current operator $\hat{J}^i_{\bm{k}}$ as the operator $\hat{O}^i$.
In the case of $\alpha =2$, the real and imaginary parts read
\begin{subequations} \label{M2_JJ}
    \begin{gather}
        \mathrm{Re} M^2_{J^iJ^j} = \pi e^2 \int_{\bm{k}} \sum_{n:\mathrm{occ}} g^{ij}_{n\bm{k}}, \label{real_M2JJ} \\
        \mathrm{Im} M^2_{J^iJ^j} = - \varepsilon_{kij} \frac{\pi e^2}{2} \int_{\bm{k}} \sum_{n:\mathrm{occ}} \Omega^k_{n\bm{k}}.
    \end{gather}
\end{subequations}
Here, $g^{ij}_{n\bm{k}}$ and $\Omega^k_{n\bm{k}}$ are the quantum metric tensor and the Berry curvature at the $n$-th band. 
This sum rule was originally discussed in Ref.~\cite{PhysRevB.62.1666}.
More recently, approaches related to this sum rule such as a partial sum rule \cite{xmz7-jgl6} and a quantum optics measurement \cite{PhysRevLett.131.156901} have been discussed to experimentally detect the quantum metric.
In two-dimensional insulating systems, the inequality
\begin{equation}
    \sqrt{\det G} \geq |\mathrm{Ch}|
\end{equation}
is produced. Here, $\mathrm{Ch} = \sum_{n:\mathrm{occ}} \int_{\mathrm{BZ}} \Omega_{n\bm{k}}/2\pi d^2 \bm{k} $ is the Chern number, which takes an integer, and $G_{ij} = \sum_{n:\mathrm{occ}} \int_{\mathrm{BZ}} g^{ij}_{n\bm{k}} /\pi d^2 \bm{k} $. This inequality is similar to Eqs.~(15) in Ref.~\cite{PhysRevB.104.045103}; however, it is slightly different with respect to the order of operations between the integration and nonlinear calculations such as square root or determinant.

In addition, we comment on a condition for equality.
The interband contribution of the tensor with the electric current operator for general $\alpha$ can be rewritten by a quantum geometric tensor, referring to Eq.~(\ref{QGT_linear_response}),
\begin{equation}
    \tilde{M}^{\alpha}_{J^i J^j}
    =\pi e^2 \int_{\bm{k}} \sum_{n : \mathrm{occ}} \braket{ \partial_{k_i} u_{n\bm{k}} | \hat{Q}_{\bm{k}} (\hat{H}_{\bm{k}} - \epsilon_{n\bm{k}} )^{2-\alpha} \hat{Q}_{\bm{k}}   | \partial_{k_j} u_{n\bm{k}} }. \label{QGT_JJ_alpha}
\end{equation}
In particular, considering an insulator, it has a finite gap, thus the eigenvalue of $\hat{H}_{\bm{k}} - \epsilon_{n\bm{k}}$ is positive as long as it is sandwiched by $\hat{Q}_{\bm{k}}$. Taking the product with a vector $\bm{v} = (1,i)$, the tensors provide
\begin{align}
    &v^*_i \tilde{M}^{\alpha}_{J^i J^j} v_j \nonumber  \\
    &= \pi e^2 \int_{\bm{k}} \sum_{n:\mathrm{occ}} || (\hat{H}_{\bm{k}} - \epsilon_{n\bm{k}} )^{1-\alpha/2} \hat{Q}_{\bm{k}} \ket{ \bar{\partial}_k u_{n\bm{k}}} ||^2 \geq 0.
\end{align}
Here, $\bar{\partial}_k = \partial_{k_x} + i \partial_{k_y}$.
This inequality is identical to the trace inequality.
The condition for equality is established when Bloch wavefunctions satisfy the holomorphicity ($\hat{Q}_{\bm{k}} \ket{\bar{\partial}_k u_{n\bm{k}} } = 0$).
These identical conditions for equality are equal to the vortexability, the condition that a wavefunction living in bands of interest remains within the same bands even after a vortex is imposed on the wavefunction by $\hat{z} = \hat{x}+ i\hat{y}$ in two-dimensional systems \cite{PhysRevB.108.205144}.
The bands satisfying these identical conditions are called the vortexable bands, providing an exact many-body ground state of short-range interactions exhibiting the fractional quantum Hall effect if the bands are flat. 
Originally, the condition has been discussed for the quantum metric and the Berry curvature (the case of $\alpha = 2$), however, we show that equivalent conditions are established for other $\alpha$ cases as well.

\section{Generalized quantum geometric tensor} \label{sec_orbital}
In this section, we will study bounds given by a generalized quantum geometric tensor (g-QGT).
The original QGT is defined by \cite{shapere1989geometric}
\begin{align}
    Q^{ij}_{\bm{k}} &= \mathrm{Tr} [ \hat{P}_{\bm{k}} (\partial_{k_i} \hat{P}_{\bm{k}}) ( \partial_{k_j} \hat{P}_{\bm{k}})] \nonumber \\
    &= g^{ij}_{\bm{k}} - \frac{i}{2} \varepsilon_{kij} \Omega_{\bm{k}}^k.
\end{align}
The real part and the imaginary part of this QGT correspond to the quantum metric tensor $g_{\bm{k}}^{ij}$ and the Berry curvature $\Omega_{\bm{k}}^{k}$, respectively. The semipositive definiteness of this QGT leads to the well-known bound between them \cite{PhysRevB.90.165139,peotta2015superfluidity}.

\subsection{A generalized QGT with semipositive definiteness and inequalities}
We introduce a g-QGT as
\begin{equation}
    R^{ij}_{\bm{k}} = \mathrm{Tr} [ \hat{O}_{\bm{k}} (\partial_{\lambda_i} \hat{P}_{{\bm{k}}}) (\partial_{\lambda_j} \hat{P}_{{\bm{k}}}) ]. 
\end{equation}
Here, $\hat{O}_{\bm{k}}$ is a Hermitian operator with semipositive definiteness, thus $\braket{v | \hat{O}_{\bm{k}} | v} \geq 0$ for any state $\ket{v}$.
No assumptions other than these two properties of $\hat{O}_{\bm{k}}$ are required in the following discussion.
$\lambda$ is a general parameter such as the Bloch wavenumber $\lambda = k$.
Then, we can prove that this g-QGT is also semipositive as
\begin{align}
    v^{*}_{i} R^{ij}_{\bm{k}} v_j = \sum_{v} \braket{v | \hat{O}_{\bm{k}} | v} \geq 0
\end{align}
for any complex vector $\bm{v}$.
After taking a sum over $\bm{k}$ in the B.Z., the integrated tensor $R^{ij} (\equiv \int_{\bm{k}} R^{ij}_{\bm{k}})$ is also semipositive in the same manner.
Then, if we consider the $2 \times 2$ matrix $R$, we obtain an inequality between real and imaginary parts as
\begin{equation}
    \frac{1}{2} \mathrm{tr} [\mathrm{Re} R] \geq \sqrt{\det[ \mathrm{Re} R]} \geq | \mathrm{Im} R^{12}|. \label{ineq_R}
\end{equation}

If we choose the projection operator $\hat{P}_{\bm{k}}$ as $\hat{O}_{\bm{k}}$, this g-QGT is reduced to $Q^{ij}_{\bm{k}}$.
In the following, we will discuss another operator as $\hat{O}_{\bm{k}}$.

\subsection{Interband Drude weight and orbital magnetization}
We set $\hat{O}_{\bm{k}} = |\hat{H}_{\bm{k}} - \mu| \equiv (\hat{H}_{\bm{k}} - \mu)(1 - 2 \hat{P}_{\bm{k}})$ and $\lambda = k$ \cite{resta2017,Resta_2018}.
This operator is semipositive by definition.
The real part and imaginary part of the g-QGT given by this operator are
\begin{subequations} \label{drude_om}
    \begin{gather}
        \mathrm{Re} R^{ij}_{\bm{k}}
        =
        \sum_{n: \mathrm{occ}} \mathrm{Re} \braket{ \partial_{k_i} u_{n\bm{k}} | \hat{H}_{\bm{k}} - \epsilon_{n\bm{k}} | \partial_{k_j} u_{n\bm{k}} }, \\
        \mathrm{Im} R^{ij}_{\bm{k}}
        =
        \sum_{n: \mathrm{occ}} \mathrm{Im} \braket{ \partial_{k_i} u_{n\bm{k}} | 2\mu - \hat{H}_{\bm{k}} - \epsilon_{n\bm{k}} | \partial_{k_j} u_{n\bm{k}}}.
    \end{gather}
\end{subequations}
The real part corresponds to the Eq.~(\ref{QGT_JJ_alpha}) for $\alpha = 1$.
A physical interpretation can be assigned to these parts \cite{resta2017,Resta_2018}.
The real part is related to the Drude weight $D_{ij}$.
The Drude weight can be decomposed into two terms as
\begin{equation} \label{def_drude}
    D_{ij} = e^2 \sum_{n :\mathrm{occ}} \int_{\bm{k}} \frac{\partial^2 \epsilon_{n\bm{k}}}{\partial k_i \partial k_j} = D_{ij}^{\mathrm{intra}} - D^{\mathrm{inter}}_{ij},
\end{equation}
where 
the first term ($D^{\mathrm{intra}}_{ij} = e^2 \sum_{n:\mathrm{occ}} \int_{\bm{k}} \braket{ u_{n\bm{k}}  | \frac{\partial^2 \hat{H}_{\bm{k}}  }{ \partial k_i \partial k_j }  | u_{n\bm{k}}  }$) is an intraband Drude weight, which is finite even for electron gas systems and single-band systems.
For the first-principle Hamiltonian ($\hat{H} = \hat{\bm{p}}^2/2m_e + V(\hat{\bm{x}})$, $V(\hat{\bm{x}})$ is a periodic potential), $D^{\mathrm{intra}}_{ij} = e^2 n_e \delta_{ij} /m_e$.
On the other hand,
the second term ($D^{\mathrm{inter}}_{ij} = 2 e^2 \sum_{n:\mathrm{occ}} \int_{\bm{k}} \mathrm{Re} \braket{ \partial_{k_i} u_{n\bm{k}} | \hat{H}_{\bm{k}} - \epsilon_{n\bm{k}} | \partial_{k_j} u_{n\bm{k}} }$) is an interband Drude weight, which is finite only in multiband systems and unique to periodic crystals. The latter corresponds to the real part of the tensor $R_{ij}$: $D_{ij}^{\mathrm{inter}} = 2e^2 \mathrm{Re} R^{ij}$. 
In insulators or flatband cases, two Drude weights are identical ($D_{ij}^{\mathrm{intra}} = D_{ij}^{\mathrm{inter}}$), i.e., cancel with each other because the total Drude weight $D_{ij}$ vanishes in Eq.~(\ref{def_drude}).
The imaginary part is related to the orbital magnetization as
\begin{equation}
    M^i_{\mathrm{orb}} = -\frac{e}{2} \varepsilon_{ijk} \mathrm{Im} R^{jk}. \label{eq_orbital}
\end{equation}

Then, we obtain inequalities between the Drude weight and the orbital magnetization as
\begin{equation}
    \frac{1}{2e} \sqrt{\det[ D^{\mathrm{inter}} ]} \geq |M_{\mathrm{orb}}|. \label{ineq_drude_om}
\end{equation}
While the discussion thus far has focused on absolute zero temperature, this inequality remains valid even at finite temperatures 
and it is derived in combination with the inequality that follows from the concavity of a grand potential density.
(see Appendix \ref{DW_orbital_finite_temp} for details).
For insulators with the first-principle Hamiltonian, the inequality becomes
\begin{equation}
    n_e \mu_{\mathrm{B}} \geq |M_{\mathrm{orb}}|, \label{ineq_density_om}
\end{equation}
where $\mu_{\mathrm{B}} = e \hbar / 2 m_e$ is the Bohr magneton.
In ordinary metals, the Drude weight $D_{ij}$ is positive, then, the inequality in Eq.~(\ref{ineq_density_om}) is also valid even for metals.
If focusing on particular bands of interest, e.g., using a tight binding model near a Fermi surface, the inequality in Eq.~(\ref{ineq_drude_om}) should be employed.

\begin{figure*}[t]
\includegraphics[width=0.45\linewidth]{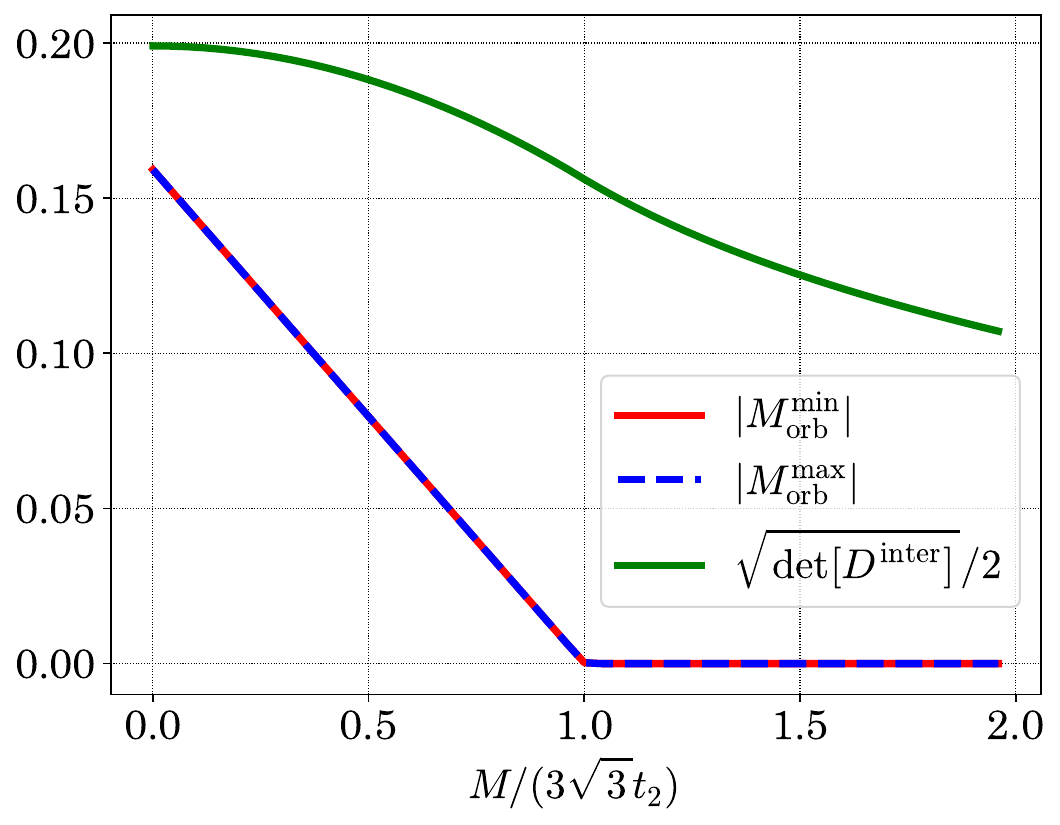}
\includegraphics[width=0.45\linewidth]{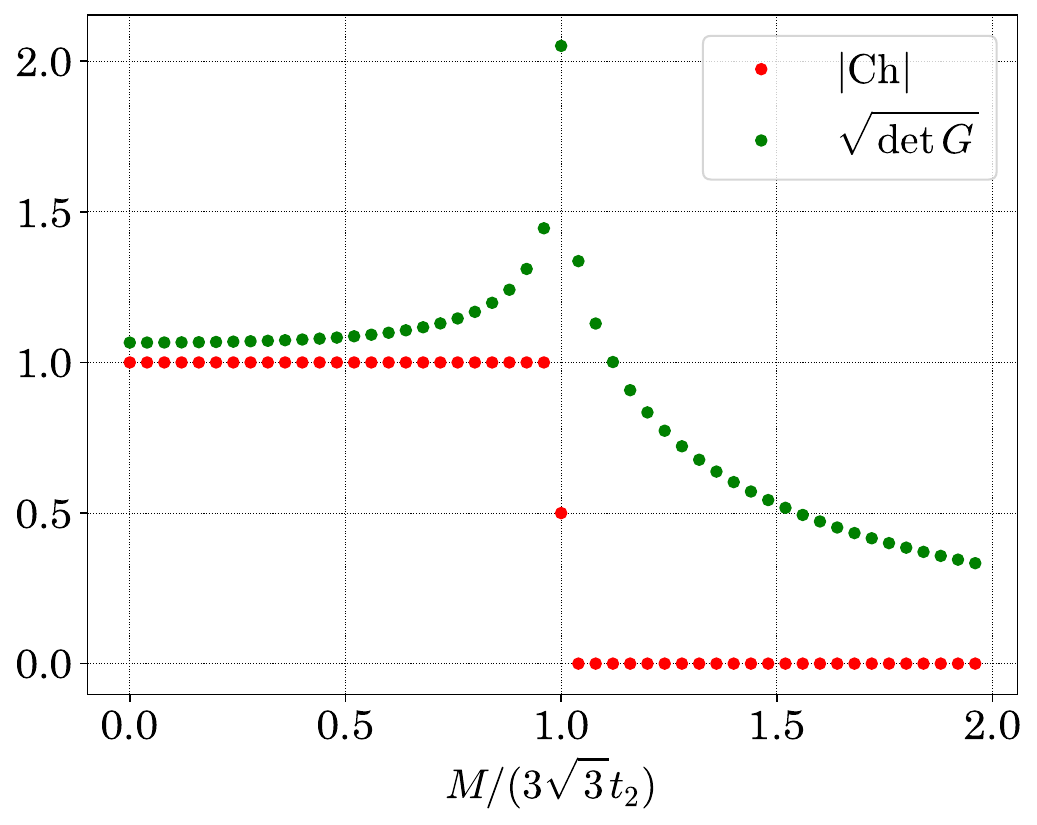}
\caption{(Left) The $M$ dependence of the interband Drude weight and the orbital magnetization in Haldane model.
(Right) The $M$ dependence of the Chern number (Ch) and $\sqrt{\det G}$.
We set $t_1 = 1.0$, $t_2 = 1/3\sqrt{3}$ for the numerical calculation.
We set $M_{\mathrm{orb}}$ in units of $et_1/\hbar$.
} \label{fig_haldane}
\end{figure*}

\subsubsection{Case study: Landau level system}
The Landau level system is described as a two-dimensional free electron gas under an out-of-plane magnetic field.
In this system, the quantum metric and the Berry curvature locally satisfy the condition of equality at all $\bm{k}$-points \cite{PhysRevB.104.045103}.

Here, we assume that the system is insulating and the number of occupied bands is set to $r$. The interband Drude weight and the orbital magnetization are given by (see Appendix \ref{appendix_landau} for details), setting $e=1$ in the following,
\begin{subequations}
    \begin{gather}
        D^{\mathrm{inter}}_{ij} =  \frac{r|B|}{2\pi m_e} \delta_{ij}, \\
        M_{\mathrm{orb}} =  \frac{r \mu}{2\pi} \mathrm{sign}(B) - \frac{r^2 B}{2 \pi m_e}.
    \end{gather}
\end{subequations}
Here, $\mu$ is the chemical potential and it lies within the gap: $\epsilon_{r-1} \leq \mu \leq \epsilon_{r} = |B|(r + 1/2)/ m_e $.
$B$ is the magnetic field. $\mathrm{sign}()$ is the sign function.
$\mu$-dependence indicates the contribution from the edge in topological insulators. Thus, the imaginary part moves within the range,
\begin{equation}
    |M_{\mathrm{orb}}| \leq \frac{r|B|}{4 \pi m_e} = \frac{1}{2} \sqrt{\det[D^{\mathrm{inter}}]} = n_e \mu_{\mathrm{B}}.
\end{equation}
In the Landau level system, the number density is $n_e = r |B|/ 2 \pi$.
We reconfirm that the inequality in Eq.~(\ref{ineq_drude_om}) holds and find that the equality holds when the edge modes are either fully occupied ($\mu = \epsilon_{r}$) or completely empty ($\mu = \epsilon_{r-1}$).

\subsubsection{Case study: Haldane model}
Haldane model is a tight-binding model with a ferromagnetic loop current order described by imaginary hoppings \cite{PhysRevLett.61.2015}.
The Hamiltonian is given by
\begin{equation}
    H_{\bm{k}}
    =
    \begin{pmatrix}
        2t_2 \sum_j \sin( \bm{k} \cdot \bm{a}'_j) + M & t_1 \sum_j e^{-i\bm{k} \cdot \bm{a}_j} \\
        t_1 \sum_j e^{i\bm{k} \cdot \bm{a}_j} & 2t_2 \sum_j \sin( \bm{k} \cdot \bm{a}'_j) - M
    \end{pmatrix},
\end{equation}
where $t_{1(2)}$ is a (next) nearest-neighbor hopping and $\bm{a}_j^{(\prime)}~(j=1,2,3)$ are vectors connecting (next) nearest-neighbor sites. $M$ is an energy difference at different two sublattices.
This model has two phases; topologically trivial ($\mathrm{Ch} = 0$) and nontrivial ($\mathrm{Ch} = 1$) phases, as shown in the right panel in FIG.~\ref{fig_haldane}.
The transition point is at $M/(3\sqrt{3}t_2) = 1$.
We can check that the inequality between the Chern number (Ch) and $\sqrt{\det G}$ in both phases is valid.
The left panel plots the $M$-dependence of the absolute value of orbital magnetization and $\sqrt{\det D^{\mathrm{inter}}}/2$.
We note that the orbital magnetization linearly depends on the chemical potential, originating from the edge modes, thus we plot it exclusively for the cases where the edge mode is completely empty ($M^{\mathrm{min}}_{\mathrm{orb}}$) and completely filled ($M^{\mathrm{max}}_{\mathrm{orb}}$).
Generally, in insulators, the orbital magnetization lies between these two values. We can confirm the inequality in Eq.~(\ref{ineq_drude_om}) and this inequality is weaker than the above one.

\subsubsection{Case study: flat band model}
\begin{figure*}[t]
\includegraphics[width=0.45\linewidth]{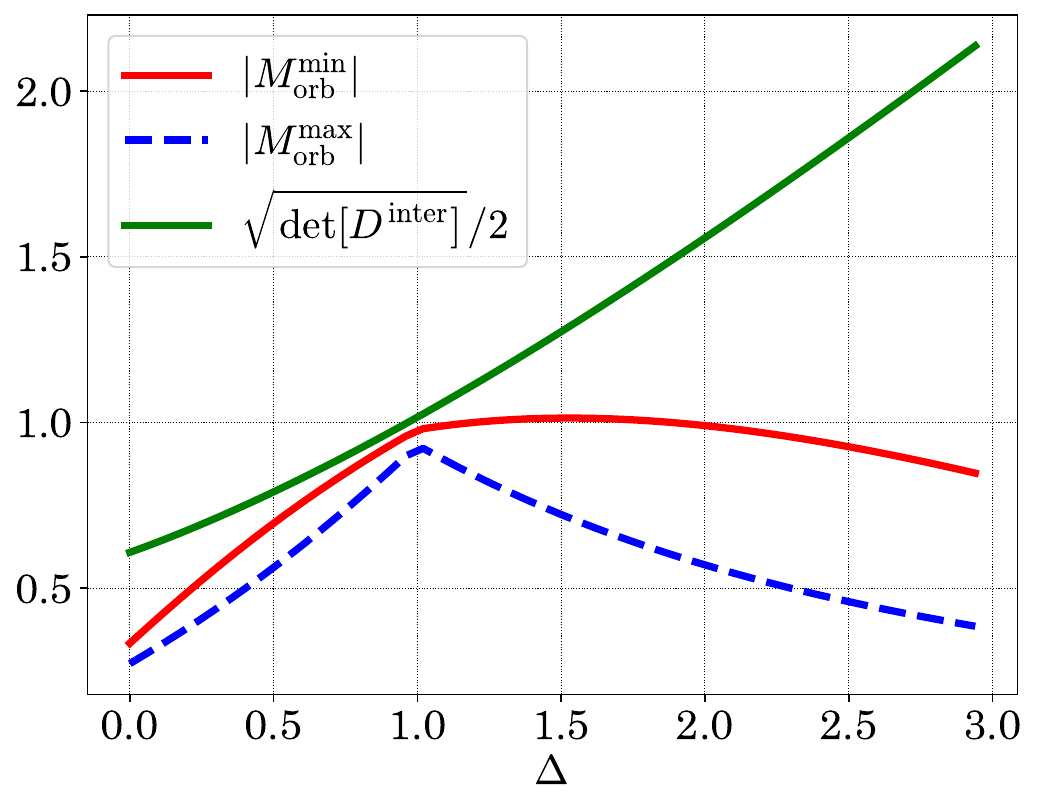}
\includegraphics[width=0.45\linewidth]{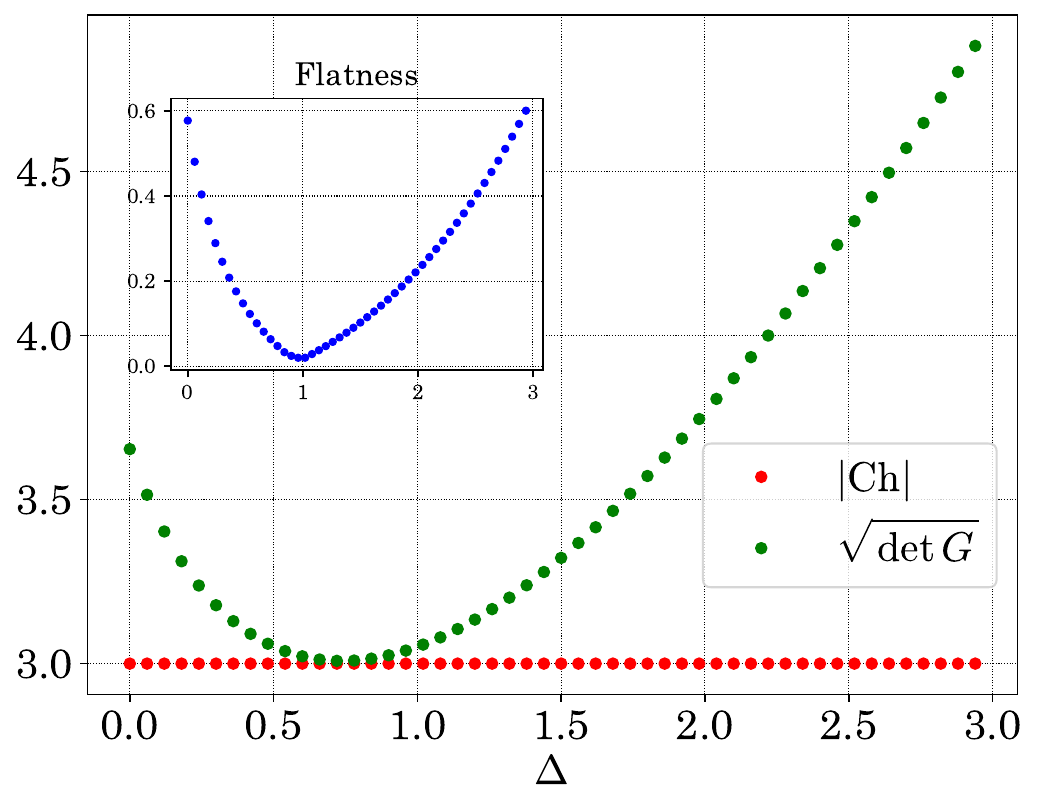}
\caption{(Left) The $\Delta$ dependence of the interband Drude weight and the orbital magnetization in the flat band model.
(Right) The $\Delta$ dependence of the Chern number (Ch) and $\sqrt{\det G}$. 
(Inset) The $\Delta$ dependence of the flatness.
We set $t_1=1.0,~\phi=\pi/3$, and $t_2 = -\Delta/\sqrt{3}$ for the numerical calculation.
We set $M_{\mathrm{orb}}$ in units of $et_1/\hbar$.
}
\label{fig_flat}
\end{figure*}
The flatness of bands is empirically essential to the condition for equality.
Actually, the inequality between the quantum metric and the Berry curvature is saturated when bands become flatter \cite{PhysRevB.104.045103}. We now examine how the inequality between the interband Drude weight and the orbital magnetization behaves in that case.
We use the model that can tune the flatness of bands, and the Hamiltonian \cite{PhysRevB.86.241112} is
\begin{equation}
    H_{\bm{k}}
    =
    \begin{pmatrix}
        \mathcal{F}(\phi)/2 & \mathcal{G}_2(5\phi) & \mathcal{G}_1(6\phi) \\
        \mathcal{G}_1(2\phi) & \mathcal{F}(3\phi)/2 & \mathcal{G}_2(7\phi) \\
        \mathcal{G}_2(3\phi) & \mathcal{G}_1(4\phi) & \mathcal{F}(5\phi)/2 \\
    \end{pmatrix} + \mathrm{h.c.},
\end{equation}
where $\mathcal{F}(\phi) = 2 t_2 \cos (k_x + k_y - \phi)$, $\mathcal{G}_1(\phi) = t_1 ( e^{ik_x} + e^{-ik_y + i\phi}) $, and $\mathcal{G}_2(\phi) = t_2 e^{i k_x - ik_y +i \phi}$.
This model is a three-orbital model defined on a square lattice, incorporating both nearest-neighbor and next-nearest-neighbor hoppings. 
The hoppings are complex with phase twists $\phi$.
We use $t_1 =1$, $\phi=\pi/3$, and $t_2 = -\Delta/\sqrt{3}$.
We can tune the flatness by changing the value of $\Delta$.
The flatness of a band is defined as the ratio between the band width and the band gap. 
We set the chemical potential within the gap between the lowest energy band and the second lowest band.
The flatness is shown in the inset in FIG.~\ref{fig_flat}, the band becomes the most flat at $\Delta \sim 1$.
Around there, the inequalities are almost saturated.

\subsubsection{Real materials}
Moir\'e Chern insulator is a flat band system with orbital ferromagnetism.
An experiment measures the orbital magnetization in the twisted bilayer graphene (TBG), and $M_{\mathrm{orb}} \approx 3.08 \times 10^{-2} \mu_{\mathrm{B}}/\mathrm{nm}^{2}$ in a Chern insulator regime (carrier density: $n_c = 2.36 \times 10^{12}~\mathrm{cm}^{-2}$, corresponding to the filling factor $\nu = +3$) \cite{tschirhart2021imaging}.
This TBG possesses eight isolated bands, with the filling factor ranging from $\nu =-4$ to $\nu = +4$ \cite{doi:10.1126/science.aay5533}.
In the case of $\nu = + 3$, the number density of these bands is $n_e = 5.5 \times 10^{-12}~\mathrm{cm}^{-2}$.
Using this number density, we get $n_e \mu_{\mathrm{B}} = 5.5 \times 10^{-2} \mu_{\mathrm{B}}/\mathrm{nm}^{2}$ and the inequality in Eq.~(\ref{ineq_density_om}) is valid.
The value of the orbital magnetization reaches $56\%$ of $n_e \mu_{\mathrm{B}}$, which is close to the condition for equality.
Strictly speaking, since we are focusing on specific bands, $n_e \mu_{\mathrm{B}}$ should not be used directly for comparison, and the effect of the effective mass should be taken into account. However, these bands are nearly flat, where the effective mass is larger than the bare electron mass $m_e$, leading to a reduced effective Bohr magneton and a correspondingly smaller $n_e \mu_{\mathrm{B}}$. Therefore, the inequality is expected to become tighter.
This provides supporting evidence that the system approximates the Landau level system.
Another candidate with the orbital magnetization is the kagome metals $A\mathrm{V_3Sb_5}~(A=\mathrm{Cs},\mathrm{Rb}, \mathrm{K})$.
Ref.~\cite{doi:10.1073/pnas.2303476121} calculates the orbital magnetization within the tight-binding model incorporating the $\mathrm{B_{3g}}$ orbitals. We can estimate the ratio $M_{\mathrm{orb}}/n_e\mu_{\mathrm{B}}$ to be approximately $0.1\%$.

\subsection{Electromagnetic response and magnetic quadrupole}
We set $\hat{O}_{\bm{k}} = |\hat{H}_{\bm{k}} - \mu|$, and we choose a Bloch wavenumber $k$ and a magnetic field $B$ as parameters $\lambda_{1,2}$, respectively. With respect to the diagonal parts of the real part of the tensor $\mathrm{Re} R_{ij}$, $\mathrm{Re} R^{11}$ and $\mathrm{Re} R^{22}$ correspond to the interband Drude weight $D^{\mathrm{inter}}$ and the interband term of the static spin susceptibility $\chi^{\mathrm{m}}$.
The off-diagonal part $\mathrm{Re} R^{12}$ is the spectral summation of the interband transition of the $\mathcal{T}$-even optical activity. This part vanishes for insulators due to the sum rule for the optical activity \cite{10.21468/SciPostPhys.14.5.118}. 
The imaginary part corresponds to the thermodynamic spin magnetic quadrupole moment \cite{PhysRevB.99.024404},
\begin{align}
\mathrm{Im} R^{kh} &= \int_{\bm{k}} \sum_{n:\mathrm{occ}} \mathrm{Im} \braket{ \partial_{k} u_{n\bm{k}} | 2\mu - \hat{H}_{\bm{k}} - \epsilon_{n\bm{k}} | \partial_{B} u_{n\bm{k}}} \nonumber \\
&\propto - \partial \Omega/\partial [ \partial B].
\end{align}
Here, $Q^{\mathrm{m}} = - \partial \Omega/\partial [ \partial B]$ is the thermodynamic spin magnetic quadrupole moment, defined by the derivative of the grand potential $\Omega$ by the gradient of the magnetic field $\partial B$.
An inequality also holds between these quantities.

\subsection{Relation with uncertainty principle}
The inequality derived from the semipositive definiteness of the tensor $R^{ij}_{\bm{k}}$ is mathematically equivalent to the uncertainty principle.
This tensor is regarded as an expectation value with respect to the probability-like function $\hat{O}_{\bm{k}}$, and is expressed as follows:
\begin{equation}
    R^{ij} \equiv \braket{ \hat{P}_i  \hat{P}_j }_O.
\end{equation}
For clarity, we omit $\bm{k}$ and we use $\hat{P}_i = \partial_{\lambda_i} \hat{P}$. Regarding the real part, the diagonal part is a standard deviation of $\partial_{\lambda_i} \hat{P}$, and the off-diagonal part is the expectation value of the symmetric part as
\begin{equation}
    \mathrm{Re} R^{ii} = \braket{ (\hat{P}_i)^2 }_O,~
    \mathrm{Re} R^{12} = \frac{1}{2} \braket{ \{ \hat{P}_1, \hat{P}_2 \} }_O.
\end{equation}
The imaginary part represents the non-commutativity,
\begin{equation}
    \mathrm{Im} R^{12} = \frac{1}{2i} \braket{ [\hat{P}_1, \hat{P}_2 ] }_O. \label{imag_g_QGT}
\end{equation}
Here, we define the commutator $[\hat{A},\hat{B}] = \hat{A} \hat{B} - \hat{B} \hat{A}$ and the anti-commutator $\{\hat{A},\hat{B} \} = \hat{A} \hat{B} + \hat{B} \hat{A}$.
Thus, the inequality in Eq.~(\ref{ineq_R}) leads to the Schr\"{o}dinger-type uncertainty principle,
\begin{equation}
    \braket{ (\hat{P}_1)^2 }_O \braket{ ( \hat{P}_2)^2 }_O - \frac{1}{4} \braket{ \{ \hat{P}_1, \hat{P}_2 \} }^2_O 
    \geq
    \frac{1}{4} |\braket{ [\hat{P}_1, \hat{P}_2 ] }_O|^2. \label{Shcrodinger}
\end{equation}
This demonstrates that the imaginary part of the g-QGT in Eq.~(\ref{imag_g_QGT}) stems from the non-commutativity of the operators and cannot be finite in classical systems, becoming finite solely as a result of quantum effects.
For example, the fact that orbital magnetization vanishes in thermal equilibrium in classical systems is known as the Bohr-van Leeuwen theorem, and this is consistent with the necessity of non-commutativity for orbital magnetization because it is the imaginary part of the g-QGT in Eq.~(\ref{eq_orbital}).
Moreover, it is precisely the presence of this non-commutativity that ensures the left-hand side in Eq.~(\ref{Shcrodinger}) is bounded to a finite value, making this inequality a manifestation of quantum effects.

\section{Conclusion}
In this paper, we have discussed various bounds between physical quantities,
based on the generalized quantum geometric tensors (QGTs).
First, 
we have demonstrated that 
the sum rule of the absorptive part of the linear response function gives the QGT extended to many-body systems, finite temperature, and general parameter space, and we have provided the exact proof that this QGT satisfies the semipositive definiteness, one of the general properties which the linear response function should follow.
Furthermore, we have shown that bounds exist for all linear response functions using this property.
We have presented several examples, and we have shown that one of the resulting inequalities can be tighter than the conventional thermodynamic bounds.

Second, we have found the bound between the Drude weight and orbital magnetization, using the known QGT generalizing the notion of the projection operator, and we have performed model calculations to examine in detail the bound.
In the Landau level system, the equality holds, similarly to the metric-curvature inequality. 
However, it should be noted that due to the contribution from edge states, the orbital magnetization can exhibit dependence on the chemical potential even within the bulk gap. 
The equality holds when the edge states are either fully occupied or completely empty.
Furthermore, by analyzing models in which the flatness of the band can be tuned, we found that as the band becomes flatter, the inequality approaches equality.
Importantly, the obtained inequality can be evaluated using only simple and static physical quantities, the electron density $n_e$ and orbital magnetization $M_{\mathrm{orb}}$, and they can be observed experimentally, as demonstrated in Ref.~\cite{tschirhart2021imaging}.
While the inequality is not exact when focusing on low-energy bands near the Fermi surface, in such cases Eq.~(\ref{ineq_drude_om}) should be used. 
Nonetheless, it is expected to capture the characteristic features of the electronic states in orbital ferromagnets.
Indeed, we have evaluated the inequality for a TBG and a kagome metal, and found that in the TBG case, it is very close to the equality. 
This supports the view that the electronic states in TBG are close to those of the Landau level system. 
It would be interesting to apply this analysis to other materials with orbital magnetism, such as rhombohedral graphene or twisted $\mathrm{MoTe_2}$.
In addition, we have shown that the similar inequality also holds for the spin magnetic quadrupole.
Finally, we have discussed the analogy between the QGT and the uncertainty principle. 
The imaginary part of the QGT becomes finite due to operator non-commutativity, while it vanishes in the classical limit. 
Moreover, the fact that the nontrivial inequalities are bounded to finite values can also be traced to this mechanism, thereby explicitly demonstrating that they are manifestations of quantum effects.

\section*{Acknowledgments}
K.S. acknowledges financial support from the RIKEN Special Postdoctoral Researcher (SPDR) Program.
N.N. was supported by JSPS KAKENHI Grant Numbers 24H00197, 24H02231, and 24K00583.
N.N. was supported by the RIKEN TRIP initiative.

\section*{Data availability}
The data that support the findings of this article are available from the authors upon reasonable request.

\appendix
\begin{widetext}
\section{Kramers-Kronig relation and sum rule} \label{appen_KK}
The linear response function
\begin{equation}
    \Phi_{O^i O^j}(\bm{q}, \omega)
    =
    -\frac{1}{V} \sum_{nm} \frac{\rho_{nm}}{\tilde{\omega} + E_{nm}} O^i_{nm}(\bm{q}) O^j_{mn}(-\bm{q})
\end{equation}
is analytic in the upper half of the complex $\omega$-plane ($\mathrm{Im} \omega \geq 0$),
and the Kramers-Kronig relation holds.
The Kramers-Kronig relation connects the real and imaginary parts of the linear response function,
\begin{subequations}
    \begin{align}
        &\mathrm{Re}\Phi_{O^i O^j}(\bm{q}, \omega) = \frac{1}{\pi} \int^{\infty}_{-\infty} \mathscr{P} \frac{\mathrm{Im}\Phi_{O^i O^j}(\bm{q}, \omega')}{\omega' - \omega} d\omega', \label{KK_real} \\
        &\mathrm{Im}\Phi_{O^i O^j}(\bm{q}, \omega) = \frac{-1}{\pi} \int^{\infty}_{-\infty} \mathscr{P} \frac{\mathrm{Re}\Phi_{O^i O^j}(\bm{q}, \omega')}{\omega' - \omega} d\omega' \label{KK_imag}.
    \end{align}
\end{subequations}
In the following, we focus only on the even part of $\Phi_{O^i O^j}(\bm{q}, \omega)$ for the wavenumber $\bm{q}$, since our primary interest lies in the $\bm{q} \to 0$ limit.
We denote this part with a superscript $(\mathrm{even})$.
The linear response function has a symmetry
\begin{equation}
    \Phi^{*}_{O^iO^j}(\bm{q},\omega) = \Phi_{O^iO^j}(-\bm{q},-\omega).
\end{equation}
Thus, for the even part $\Phi^{\mathrm{(even)}}_{O^iO^j}(\bm{q},\omega)$, the real part and the imaginary part are even and odd functions of $\omega$, respectively.
Therefore, using Eq.~(\ref{KK_real}) at the $\omega \to 0$ limit, we obtain
\begin{equation}
    \mathrm{Re}\Phi^{\mathrm{(even)}}_{O^i O^j}(\bm{q}, 0) = \frac{2}{\pi} \int^{\infty}_{0} \mathscr{P} \frac{\mathrm{Im}\Phi^{\mathrm{(even)}}_{O^i O^j}(\bm{q}, \omega')}{\omega'} d\omega'.
\end{equation}
Here, we use the fact that the integrand is an even function of $\omega$.
Taking the limit $\bm{q} \to 0$ of this expression yields exactly Eq.~(\ref{M1_OO_static}), and the sum rule yields the static susceptibility.
On the other hand, the sum rule for the real part of the response function does not give the static susceptibility, due to the opposite parity in $\omega$.

\section{Inequality between interband Drude weight and orbital magnetization at finite temperature} \label{DW_orbital_finite_temp}
\subsection{Mathematical preliminary: Inequality satisfied by convex and concave functions}
Considering a convex function $F(x)$ satisfying $F''(x) \geq 0$, Taylor's theorem yields
\begin{equation}
    F(x) = F(y) + F'(y)(x-y) + \int^x_y (x-t) F''(t) dt.
\end{equation}
Using this equation, we obtain
\begin{equation}
    ( F(x) - F(y) ) - (x-y)\frac{F'(x) + F'(y)}{2} = \int^x_y (x-t) F''(t) dt - (x-y)\frac{F'(x) - F'(y)}{2}. \label{C2}
\end{equation}
Integrating by parts the integral term, we obtain
\begin{align}
    \int^x_y (x-t) F''(t) dt = - (x-y) F'(y) + \int^x_y F'(t) dt.
\end{align}
The convexity $F''(x)$ means that the function $F'(x)$ is non-decreasing, $F'(a) \geq F'(b)$ for $a \geq b$. Thus, we obtain an inequality
\begin{equation}
    \int^x_y F'(t) \leq (x-y)F'(x).
\end{equation}
Using this inequality and Eq.~(\ref{C2}) yields
\begin{equation}
    ( F(x) - F(y) ) - (x-y)\frac{F'(x) + F'(y)}{2}
    \leq
    (x-y)\frac{F'(x) - F'(y)}{2}.
\end{equation}
For a concave function $G(x)~(G''(x) \leq 0)$, the corresponding inequality holds with the opposite sign, 
\begin{equation}
    ( G(x) - G(y) ) - (x-y)\frac{G'(x) + G'(y)}{2}
    \geq
    (x-y)\frac{G'(x) - G'(y)}{2}, \label{C6}
\end{equation}
that can be proven by replacing $F(x)$ with $-G(x)$.

\subsection{Interband Drude weight and orbital magnetization at finite temperature}
Here, we consider a Hermitian tensor
\begin{equation}
    \tilde{R}_{ij} = e \int_{\bm{k}} \sum_{n \neq m} \Bigl( g_{nm\bm{k}} - \frac{\tilde{f}_{nm\bm{k}} \epsilon_{nm\bm{k}} }{2} - \frac{f_{nm\bm{k}} \epsilon_{nm\bm{k}} }{2}  \Bigr) \braket{ \partial_{k_i} u_{n\bm{k}} | u_{m\bm{k}} } \braket{  u_{m\bm{k}} | \partial_{k_j} u_{n\bm{k}} }.
\end{equation}
$f_{n\bm{k}} = 1/(1 + e^{\beta( \epsilon_{n\bm{k}} - \mu)})$ is the Fermi distribution function and $g_{n\bm{k}} = - \beta^{-1} \ln ( 1 + e^{- \beta( \epsilon_{n\bm{k}} - \mu )} )$ is the grand potential density.
We use some abbreviations: $a_{nm\bm{k}} = a_{n\bm{k}} - a_{m\bm{k}}$ and $\tilde{a}_{nm\bm{k}} = a_{n\bm{k}} + a_{m\bm{k}}$.
The grand potential density $g_{n\bm{k}}$ is the concave function for $\epsilon_{n\bm{k}}$ and $\partial g_{n\bm{k}} / \partial \epsilon_{n\bm{k}} = f_{n\bm{k}}$ is satisfied.
Therefore, Eq.~(\ref{C6}) provides an inequality
\begin{equation}
    g_{nm\bm{k}} - \frac{\tilde{f}_{nm\bm{k}} \epsilon_{nm\bm{k}} }{2} - \frac{f_{nm\bm{k}} \epsilon_{nm\bm{k}} }{2} \geq 0.
\end{equation}
This inequality leads to the semipositive definiteness, as, for any complex vectors $\bm{v}$,
\begin{equation}
    v_i^* \tilde{R}_{ij} v_j
    = e \int_{\bm{k}} \sum_{n \neq m} \Bigl( g_{nm\bm{k}} - \frac{\tilde{f}_{nm\bm{k}} \epsilon_{nm\bm{k}} }{2} - \frac{f_{nm\bm{k}} \epsilon_{nm\bm{k}} }{2}  \Bigr)
    | v_i^* \braket{ \partial_{k_i} u_{n\bm{k}} | u_{m\bm{k}} } |^2 \geq 0.
\end{equation}
Then we obtain the inequality between the real part and the imaginary part of the $2 \times 2$ matrix $\tilde{R}$,
\begin{equation}
    \sqrt{\det [\mathrm{Re} \tilde{R} ] } \geq | \mathrm{Im} \tilde{R}^{12} |.
\end{equation}
The real part is the interband Drude weight
\begin{align}
    \mathrm{Re} \tilde{R}_{ij} &= -e \int_{\bm{k}} \sum_{n \neq m}
    \frac{f_{nm\bm{k}} \epsilon_{nm\bm{k}} }{2} \mathrm{Re} [\braket{ \partial_{k_i} u_{n\bm{k}} | u_{m\bm{k}} } \braket{  u_{m\bm{k}} | \partial_{k_j} u_{n\bm{k}} }] \nonumber \\
    &=
    e \int_{\bm{k}} \sum_{n}
    f_{n\bm{k}} \mathrm{Re} [\braket{ \partial_{k_i} u_{n\bm{k}} | \hat{H}_{\bm{k}} - \epsilon_{n\bm{k}} | \partial_{k_j} u_{n\bm{k}} }]
    = D^{\mathrm{inter}}/2e,
\end{align}
and the imaginary part is the orbital magnetization \cite{PhysRevLett.99.197202}
\begin{align}
    \mathrm{Im} \tilde{R}_{12}
    &=
    e \int_{\bm{k}} \sum_{n \neq m} \Bigl( g_{nm\bm{k}} - \frac{\tilde{f}_{nm\bm{k}} \epsilon_{nm\bm{k}} }{2}  \Bigr) 
    \mathrm{Im} [
    \braket{ \partial_{k_1} u_{n\bm{k}} | u_{m\bm{k}} } \braket{  u_{m\bm{k}} | \partial_{k_2} u_{n\bm{k}} } ] \nonumber \\
    &=
    \int_{\bm{k}} \sum_n ( f_{n\bm{k}} m_{n\bm{k}} - e g_{n\bm{k}} \Omega_{n\bm{k}}  ) = M_{\mathrm{orb}}.
\end{align}
Here, $m_{n\bm{k}} = e \mathrm{Im}[ \braket{ \partial_{k_1} u_{n\bm{k}} | \hat{H}_{\bm{k}} - \epsilon_{n\bm{k}} | \partial_{k_2} u_{n\bm{k}} }] $ is the orbital magnetic moment and $\Omega_{n\bm{k}} = -2 \mathrm{Im} [ \braket{ \partial_{k_1} u_{n\bm{k}} | \partial_{k_2} u_{n\bm{k}}  } ]$ is the Berry curvature.
Using the above relations, we finally obtain the inequality
\begin{equation}
    \frac{1}{2e} \sqrt{\det[ D^{\mathrm{inter}} ]} \geq |M_{\mathrm{orb}}|,
\end{equation}
which is the extension of Eq.~(\ref{ineq_drude_om}) to the finite temperature case.

\section{Interband Drude weight and orbital magnetization in the Landau level system} \label{appendix_landau}
Following Ref.~\cite{PhysRevB.104.045103}, we calculate the interband Drude weight $D^{\mathrm{inter}}$ and the orbital magnetization $M_{\mathrm{orb}}$ in the Landau level system.
The Landau level system is the free electron gas system under a uniform out-of-plane magnetic field, and the Hamiltonian reads
\begin{equation}
    \hat{H} = \frac{1}{2m_e} ( \hat{\bm{p}} + e \hat{\bm{A}} )^2.
\end{equation}
Here, $-e (< 0)$ is the electron charge, and we set $e=1$, following.
We choose the Landau gauge for the vector potential as $\hat{\bm{A}} = ( 0, B \hat{\bm{x}})$, where $B$ is a uniform magnetic field.
In this system, the ordinary translation symmetry is absent; however, it has a magnetic translation symmetry.
The magnetic unit cell is a rectangle with the length of $x$-direction $a_x$ and the one of $y$-direction $a_y$, satisfying $|B| a_x a_y = 2 \pi$.
The magnetic symmetry guarantees the application of the Bloch theorem, and we can define the Bloch wavefunction $\psi_{n\bm{k}}(\bm{r}) = e^{i\bm{k} \cdot \bm{r}} u_{n\bm{k}}(\bm{r})$, where $\bm{k}$ lives in the magnetic Brillouin Zone: $[-\pi/a_x , / \pi/ a_x) \otimes[- \pi/ a_y , \pi / a_y )$.
We leave the detailed calculations to Ref.~\cite{PhysRevB.104.045103} and list the necessary formulas here:
\begin{subequations}
    \begin{gather}
        \braket{\tilde{u}_{n\bm{k}} | \tilde{u}_{n\bm{k}} } = a_y l_B \sqrt{\pi} 2^n n!, \\
        \frac{\braket{\tilde{u}_{n'\bm{k}} |\partial_x \tilde{u}_{n\bm{k}} }}{\braket{\tilde{u}_{n\bm{k}} | \tilde{u}_{n\bm{k}} }} = 
        -i l_B [ (n+1) \delta_{n+1,n'} + \frac{1}{2} \delta_{n-1,n'} ] +\frac{ik_y}{B} \delta_{n,n'}, \\
        \frac{\braket{\tilde{u}_{n'\bm{k}} |\partial_y \tilde{u}_{n\bm{k}} }}{\braket{\tilde{u}_{n\bm{k}} | \tilde{u}_{n\bm{k}} }} 
        =
        - \frac{1}{l_B B} [ -(n+1) \delta_{n+1,n'} + \frac{1}{2} \delta_{n-1,n'} ].
    \end{gather}
\end{subequations}
Here, $l_B = 1/\sqrt{|B|}$ is the magnetic length and $\epsilon_{n} = |B|/m_e(n+1/2)~(n \geq 0)$ is the eigenenergy. 
First, the real part of the g-QGT can be rewritten by
\begin{align}
    \mathrm{Re} R^{ij}_{\bm{k}}
    &=
    \sum_{n: \mathrm{occ}} \mathrm{Re} \braket{ \partial_{k_i} u_{n\bm{k}} | \hat{H}_{\bm{k}} - \epsilon_{n\bm{k}} | \partial_{k_j} u_{n\bm{k}} } \nonumber \\
    &=
    \sum_{n:\mathrm{occ}} \sum_{m:\mathrm{unocc}} \epsilon_{mn\bm{k}} \mathrm{Re} \braket{ \partial_i u_{n\bm{k}} | u_{m\bm{k}} } \braket{ u_{m\bm{k}} | \partial_j u_{n\bm{k}} }.
\end{align}
Using the above formulas, the real part is
\begin{equation}
    \mathrm{Re} R^{ij}_{\bm{k}} = (\epsilon_{r\bm{k}} - \epsilon_{r-1\bm{k}})
    \mathrm{Re} \biggl( \frac{\braket{\partial_i \tilde{u}_{r-1\bm{k}} | \tilde{u}_{r\bm{k}}}  \braket{ \tilde{u}_{r\bm{k}} | \partial_j \tilde{u}_{r-1\bm{k}}}}{\braket{\tilde{u}_{r-1\bm{k}} | \tilde{u}_{r-1\bm{k}} } \braket{\tilde{u}_{r\bm{k}} | \tilde{u}_{r\bm{k}} } } \biggr).
\end{equation}
Each component is given by
\begin{subequations}
    \begin{gather}
        \mathrm{Re} R^{xx}_{\bm{k}}
        =
        \frac{|B|}{m_e} (l_B r)^2 \frac{2^{r-1} (r-1)!}{2^{r} r!} = \frac{r}{2m_e}, \\
        \mathrm{Re} R^{yy}_{\bm{k}}
        =
        \frac{|B|}{m_e} \Bigl( \frac{r}{l_B B} \Bigr)^2 \frac{2^{r-1} (r-1)!}{2^{r} r!} = \frac{r}{2m_e}, \\
        \mathrm{Re} R^{xy}_{\bm{k}} = \mathrm{Re} R^{yx}_{\bm{k}} = 0.
    \end{gather}
\end{subequations}
Second, the imaginary part of the g-QGT can be rewritten by
\begin{align}
   \mathrm{Im} R^{ij}_{\bm{k}}
    &=
    \sum_{n: \mathrm{occ}} \mathrm{Im} \braket{ \partial_{k_i} u_{n\bm{k}} | 2\mu - \hat{H}_{\bm{k}} - \epsilon_{n\bm{k}} | \partial_{k_j} u_{n\bm{k}}} \nonumber \\
    &=
    \sum_{n:\mathrm{occ}} \sum_{m:\mathrm{unocc}} (2\mu - \epsilon_{m\bm{k}} - \epsilon_{n\bm{k}}) \mathrm{Im} \braket{\partial_i u_{n\bm{k}} | u_{m\bm{k}}}  \braket{ u_{m\bm{k}} |\partial_j u_{n\bm{k}}}.
\end{align}
Using the above formulas, it can be calculated as
\begin{align}
    \mathrm{Im} R^{xy}_{\bm{k}}
    =&
    (2 \mu - \epsilon_{r\bm{k}} - \epsilon_{r-1 \bm{k}} )
    \mathrm{Im} 
    \biggl( \frac{\braket{\partial_x \tilde{u}_{r-1\bm{k}} | \tilde{u}_{r\bm{k}}}  \braket{ \tilde{u}_{r\bm{k}} | \partial_y \tilde{u}_{r-1\bm{k}}}}{\braket{\tilde{u}_{r-1\bm{k}} | \tilde{u}_{r-1\bm{k}} } \braket{\tilde{u}_{r\bm{k}} | \tilde{u}_{r\bm{k}} } } \biggr) \nonumber \\
    =&
    -\Bigl(
    2\mu - \frac{2|B|}{m_e} r
    \Bigr)
    \frac{r^2l_B}{l_B B} \frac{2^{r-1} (r-1)!}{2^{r} r!} \nonumber \\
    =&
    -\frac{r \mu}{B} + \frac{r^2}{m_e} \mathrm{sign}(B).
\end{align}
Similar to the quantum metric and the Berry curvature, these quantities are uniform across the Brillouin zone in the Landau level system.
Then, integrating these quantities over the magnetic Brillouin zone, we obtain the interband Drude weight and the orbital magnetization as
\begin{subequations}
    \begin{gather}
        D^{\mathrm{inter}}_{xx} = D^{\mathrm{inter}}_{yy} = \int_{\mathrm{B.Z.}} \frac{d^2 \bm{k}}{(2\pi)^2} 2\mathrm{Re} R^{xx(yy)}_{\bm{k}} = \frac{r |B|}{ 2 \pi m_e },
        ~~~D_{xy}^{\mathrm{inter}} = D_{yx}^{\mathrm{inter}} = 0,
        \\
        M_{\mathrm{orb}} = - \int_{\mathrm{B.Z.}} \frac{d^2 \bm{k}}{(2\pi)^2} \mathrm{Im} R^{xy}_{\bm{k}}
        =
        \frac{|B|}{2 \pi } \Bigl(  \frac{r \mu}{B} - \frac{r^2}{m_e} \mathrm{sign}(B)  \Bigr).
    \end{gather}
\end{subequations}
Here, we use the equation $|B| a_x a_y = 2 \pi$. 
In insulators, the interband Drude weight is identical to the intra Drude weight.
We can explicitly check this identity as
\begin{equation}
    D^{\mathrm{intra}}_{ij} = \sum_{n:\mathrm{occ}} \int_{\mathrm{B.Z.}} \frac{d^2 \bm{k}}{(2\pi)^2} \braket{ u_{n\bm{k}} | \frac{\partial^2 \hat{H}_{\bm{k}} }{ \partial k_i \partial k_j}  | u_{n\bm{k}} } 
    =
    \frac{n_e}{m_e} \delta_{ij}.
\end{equation}
Here, $n_e = r |B| / 2\pi$ is the number density of electrons occupying the $r$ Landau levels.

\end{widetext}

\bibliography{reference.bib}

\end{document}